\documentclass[aps,prd]{revtex4}
\usepackage{graphicx}
\begin{document}
\title{Influence of statistical fluctuations on
$K/\pi$ ratios in relativistic heavy ion collisions}
\author{C.B.
Yang$^{1, 2}$ and X. Cai$^1$}
\affiliation{$^1$ Institute of Particle
Physics, Hua-Zhong Normal University, Wuhan 430079, China\\
$^2$ Max-Planck-Institut f\"ur Physik, F\"ohringer Ring 6, D-80805
 M\"unchen, Germany}
\date{\today}

\begin{abstract}
The influence of pure statistical fluctuations on $K/\pi$ ratio
is investigated in an event-by-event way. Poisson and the modified
negative binomial distributions are used as the multiplicity
distributions since they both have statistical background.
It is shown that the distributions of the ratio in these cases are
Gaussian, and the mean and relative variance are given analytically.


\end{abstract}
\maketitle

The ultimate goal of current and future ultra-relativistic heavy ion program
is the production and characterization of an extended volume of deconfined
phase of quarks and gluons, the quark-gluon plasma (QGP). The possible existence
of QGP as the equilibrium state of strongly interacting matter has been
predicted in lattice theories \cite{jcc} at sufficiently high temperature
and density. So QGP might exist in the early universe and inside the neutron
stars. Heavy ion collision at extremely high energies is the only chance to
study QGP in a controlled way. In laboratory experiments, however, the QGP can
survive only for very short period of time, and it will cool during its
expanding and become hadronic matter at freeze-out point. It is very difficult
to detect QGP in experiments because most events are, due to collision
geometry and quantum fluctuations, without QGP even if the condition for
QGP creation can be reached in the experiments. So, a variety of possible
signatures for the transient existence of the deconfined state of matter
in the collisions have been proposed theoretically and experimentally (see
\cite{bgsg99,pr236,na638} for a review). One of the most important and most
interesting signatures is the enhanced strangeness production in the
process of heavy ion collisions \cite{pr142} compared with that in $p-p$
collisions. The investigation of strangeness production is useful for
understanding the mechanism of heavy quark production since no strangeness
content exists in the initial state in heavy-ion collisions. Strangeness
production can also shed light on the time scale of chemical freeze-out
and therefore can carry information about the early stage of a heavy ion
collision \cite{lpc94}.

Great progress has been made with the proposal and applications of
event-by-event analysis \cite{zpc54,epjc6,appb29,plb435,plb439,plb459,epjc8}
in the study of high energy heavy ion collisions. Event-by-event analysis of
heavy ion collisions became possible with the advent of large acceptance
detectors \cite{la99}. The philosophy of event-by-event physics is based on
following speculation: Although the conditions to produce QGP may be reached
in every event, the fact that a phase transition is a critical phenomenon
implies that it may occur only in a very small sub-sample of events. So the
fluctuations accompanying the phase transition will, in effect, be averaged
out in the conventional ensemble analyses. The event-by-event analysis
searches for fluctuations of observables at the event level, so it can be used
to select interesting or anomalous event candidates \cite{nato94}
with specific dynamical properties. So the new method
can provide dynamical information which cannot be obtained from
the traditional inclusive spectra.

Recently the event-by-event analysis is used in experiments to analyze
strangeness productions in high energy heavy ion collisions. In
\cite{gr97,fs99} the kaon to pion ratio for single event is studied as
the event observable with which one can look at fluctuations in
the chemical freeze-out stage of the collisions. Of course, the distribution
of the $K/\pi$ ratio is the result of a combination of statistical and
possible dynamical fluctuations in the strangeness production. The method of
mixed events, in which there are no momentum correlation among
particles by construction, is used in the experimental analyses
to estimate the effect of the former on
the distribution of $K/\pi$ ratio. Both distributions for $K/\pi$ ratio from
the real data and the mixed events are Gaussian, and the widths are only with
a little difference. That means the dominance of statistical fluctuations
in strangeness production. Although it is well established that there is no
spatial-temporal correlation in the mixed events, we do not know exactly
if there exist some other correlations. In the study of $K/\pi$ ratio the
only quantities concerned are the numbers of produced kaons and pions. The
event-mixing technique will not change the total numbers of the particles
produced. These numbers carry some important information
about the chemical fluctuations at the freeze-out point. So, some information
about the strangeness production may still be retained in the mixed events.
To investigate the problem theoretically further one should ask: What is
the distribution of $K/\pi$ ratio if there exist only statistical multiplicity
fluctuations in the collisions? Will it be a Gaussian as showed from
the mixed events? If yes, what are the mean and the width? Are they in
agreement with those obtained from the mixed events?
What will be the dependences of them on mean multiplicities of
kaons and pions?

In this paper the $K/\pi$ ratio for the case with pure statistical fluctuations
is studied in an event-by-event way with two assumptions: (1) There is no
dynamical fluctuation in the productions of both the kaons and pions,
and (2) there is no correlation between the productions of the two kinds of
particles.
This is, of course, a trivial case, but it is the base for any event-by-event
investigation on the influence of dynamical fluctuations on $K/\pi$ ratio.

Since we have assumed no dynamical fluctuations in the productions of
kaons and pions nor correlation between them, the distributions of the
multiplicities of kaons and pions can be specified from some statistical
considerations. Two possibilities will be considered in this paper.
As a first possibility, the distributions can be Poissonian,
i.e. the probabilities
with which there are $k$ kaons and/or $\pi$ pions in an event are,
respectively,
\begin{equation}
p_K={\langle k\rangle^k\over k!}{\rm e}^{-\langle k\rangle},\quad\quad
p_\Pi={\langle \pi\rangle^\pi\over \pi!}{\rm e}^{-\langle \pi\rangle}
\end{equation}

\noindent with $\langle k\rangle$ and $\langle \pi\rangle$ the mean
multiplicities of kaons and pions in the collisions. In a global
analysis, the global $K/\pi$ ratio is then
\begin{equation}
R_g=\langle k\rangle/ \langle \pi\rangle\quad.
\end{equation}

\noindent
From the view point of the theory of coherent state the assumption of
Poisson distributions for the multiplicities implies
that both kaons and pions are emitted from some coherent sources and
that no thermal effect is taken into account.
Now we turn to event-by-event analysis of the ratio.
For an event with $k$ kaons and $\pi$ pions the $K/\pi$ ratio is
\begin{equation}
r_e\equiv{k\over \pi}\ ,
\end{equation}

\noindent and the probability density for $K/\pi$ ratio to be in neighborhood
of $r_e$ can be given by
\begin{equation}
P_{r_e}=\sum_{r_e=k/\pi} p_K(k)p_\Pi(\pi)=\sum_{\pi=1}^\infty
{\langle k\rangle^{r_e\pi}\over (r_e\pi)!}{\langle\pi\rangle^\pi\over \pi!}
{\rm e}^{-(1+R_g)\langle\pi\rangle}\quad,
\label{pre}
\end{equation}

\noindent because the productions of kaons and pions have been
assumed to be independent.

Since $p_K$ and $p_\Pi$ are peaked at $\langle k\rangle$ and
$\langle \pi\rangle$ respectively, it is easy to see that $P_{r_e}$
will be peaked near $R_g$. A natural question is: Can $P_{r_e}$ be
Gaussian for given $\langle k\rangle$ and $\langle\pi \rangle$ ?
Since Eq. (\ref{pre}) is too complicated, Monte Carlo simulation can be
used to answer this question. To get an event for this case, one needs only
to generate two independent random integer numbers (according to two Poisson
distributions with means $\langle k\rangle$ and $\langle\pi\rangle$
respectively). The random integer number corresponding to $\langle k\rangle$
is taken to be the kaon multiplicity in the event, the other for pions.
Then one gets a corresponding $K/\pi$ ratio $r_e$ for the event.
The distribution of $r_e$ from the simulation is given in Fig. 1
with 1 million events used. In the simulation the mean multiplicities
for kaons and pions are chosen to be 38 and 200 respectively
for illustration. Here we choose $R_g=0.19$, the same as shown experimentally
in Pb-Pb collisions at 158 $A$ GeV, for later comparison with experimental
result. One can see that the distribution of the $K/\pi$ ratio
is indeed approximately a Gaussian. This is consistent
with the conclusions given from the mixed events in Refs. \cite{gr97,fs99}.
To get more information about the distribution one may want to know the
mean and width of the distribution. From Eq. (\ref{pre}) one can see that the
distribution of $r_e$ is not symmetric about $R_g$, so the
mean value of the distribution is, to some extent, different from $R_g$.
The mean value of the distribution for the trivial case can be calculated
analytically as
\begin{mathletters}
\begin{equation}
R_e=\langle {k\over \pi}\rangle\equiv\left.\sum_{k=0}^{\infty}
\sum_{\pi=1}^{\infty}{k\over\pi}p_K(k)p_\Pi(\pi)\right/
\sum_{k=0}^\infty\sum_{\pi=1}^\infty p_K(k)p_\Pi(\pi)
\label{a1}
\end{equation}
\begin{equation}
\simeq R_g\sum_{\pi=1}^\infty
{\langle\pi\rangle^{\pi+1}\over \pi\pi!}{\rm e}^{-\langle\pi\rangle}\ .
\mbox{\hspace*{3.5cm}}
\label{a2}
\end{equation}
\end{mathletters}

\noindent In the transition from Eq. (\ref{a1}) to Eq. (\ref{a2}) condition
$\langle\pi\rangle\gg 1$ has been used. The relative accuracy in Eq.
(\ref{a2}) is of the order of ${\rm e}^{-\langle\pi\rangle}$, thus the
approximation is quite good for $\langle\pi\rangle$ a few hundreds which is
quite normal in current high energy heavy ion collisions. By using the
generating function for the Poisson distribution
\begin{equation}
G_\Pi(x)=\sum_{\pi=0}^\infty {\langle\pi\rangle^\pi x^{\pi}\over \pi!}
{\rm e}^{-\langle\pi\rangle}=\sum_{\pi=1}^\infty x^{\pi}p_\Pi(\pi)={\rm e}
^{\langle\pi\rangle(x-1)}\quad,
\label{gen}
\end{equation}

\noindent the mean value of the distribution $R_e$ can be expressed as
\begin{equation}
R_e=R_g\int_0^1 dx \langle\pi\rangle {G_\Pi(x)-G_\Pi(0)\over x}\quad.
\end{equation}

\noindent We are particularly interested in the ratio $R$ between $R_e$ and
$R_g$, which is a measure of the deviation of the two $K/\pi$ ratios obtained
from global and event-by-event analyses, respectively. In our trivial case
\begin{equation}
R\equiv{R_e\over R_g}=\langle\pi\rangle
\int_0^1 dx {G_\Pi(x)-G_\Pi(0)\over x}\quad,
\label{ratio}
\end{equation}

\noindent which depends only on the mean multiplicity of pions.
It can be seen that $R$ will approach 1.0 when $\langle\pi
\rangle\to\infty$. In this limit the mean values of $K/\pi$ ratios from
global and event-by-event analyses are equal, as can be naively expected.
The change behavior of $R$ as a function of $\langle\pi\rangle$ is shown
in Fig. 2. $R$ is large for lower mean pion multiplicity and decreases
with $\langle\pi\rangle$. But the values of $R$ are always close to
the trivial value 1.0 within an accuracy of about 1\% in the graphed range
of the mean pion multiplicity. So one can hardly find the difference between
$R_e$ and $R_g$. Using the generating function for pion multiplicity
distribution, one  can also calculate the width of the $K/\pi$ ratio
distribution. First one can get
\begin{equation}
\left\langle\left({k\over\pi}\right)^2\right\rangle=
\langle k^2\rangle\int_0^1 dx\
{G_\Pi(x)-G_\Pi(0)\over x}\ln{1\over x} \quad,
\end{equation}

\noindent where $\langle k^2\rangle=\langle k\rangle^2+\langle k\rangle$,
since the distribution of the kaon multiplicity is assumed to be a
Poissonian. The width of the $K/\pi$ ratio distribution is then
\begin{equation}
\sigma=\sqrt{\left\langle\left({k\over\pi}\right)^2\right\rangle-\left
\langle{k\over\pi}\right\rangle^2}\quad.
\end{equation}

\noindent The relative variance is then
\begin{equation}
{\sigma\over R_e}=\sqrt{
{\left(1+{1\over R_g\langle\pi\rangle}\right)
\int_0^1dx\ \ln{1\over x}\ {G_\Pi(x)-G_\Pi(0)\over x}
\over \left[\int_0^1dx\ {G_\Pi(x)-G_\Pi(0)\over x}\right]^2 }-1}\quad.
\label{width}
\end{equation}

\noindent Different from $R$ the relative variance depends on both $\langle
\pi\rangle$ and $R_g$ (thus on $\langle k\rangle$). The behavior of the
relative variance as a function of $\langle\pi\rangle$ for given
$R_g=0.19$ is shown in Fig. 3. The value of $R_g$ is chosen for
illustration and is the same as that obtained from current experiments on
Pb-Pb collision at 158 $A$ GeV. For $\langle\pi\rangle$ from 100
to 300 the relative variance decreases quite quickly from about 28\%
to about 16.5\%. The relative variance given experimentally from the mixed
events is about 23\% at $\langle \pi\rangle$=270.13 \cite{croland}.
Since the relative
variance depends on $R_g$, different curve for the dependence of $\sigma/R_e$
on $\langle\pi\rangle$ will be obtained if another value of $R_g$ is taken as
input. For fixed $\langle \pi\rangle$, the larger $R_g$, the smaller
$\sigma/R_e$. The decrease tendency of $\sigma/R_e$ with $\langle\pi\rangle$
will be similar for all $R_g$. The decrease of the relative
width $\sigma/R_e$ with the increase of $\langle\pi\rangle$ can be
anticipated from the fact that both the distributions for kaons and pions
are of Poisson in which the relative widths of the multiplicity distributions
are inversely proportional to
the root of the mean multiplicities. In Refs. \cite{gr97,fs99} the mean
multiplicity of positively charged pions is 270.13, in the range discussed
in this paper. The experimental point is given in Fig. 3 by a solid square.
One can see that the theoretically calculated relative
width of the $K/\pi$ ratio distribution from the Poissonian multiplicity
distributions would be too small. The discrepancy of calculated
$\sigma/R_e$ from the experimental result for the same $\langle\pi\rangle$
shows the important role played by
the decoherent effect in the emission of the pions and kaons. The decoherent
effect will broaden the multiplicity distributions of kaons and pions, so that
a larger relative width of the $K/\pi$ ratio distribution can be expected.
Such thermal effect cannot be washed out in the so called mixed events.

As the second possibility for the multiplicity distribution with
pure statistical fluctuations, one can work with the modified negative
binomial (MNB) distribution \cite{mnb1,mnb2}. As a generalization of
the negative binomial distribution, the thermal effect plays an important
role in the MNB distribution. In this case, the generating function is
\begin{equation}
G_{\rm MNB}(x)=\left({1-\Delta(x-1)\over 1-r(x-1)}\right)^k\quad ,
\label{mnbd}
\end{equation}

\noindent with three parameters $\Delta, r$ and $k$. As shown in the
first paper in \cite{mnb1}, $k$ can be interpreted as the maximum
number of fireballs
or clusters in some initial states, $\Delta$ and $r$ can be related
to the rates of birth and immigration processes. The mean multiplicity
in the MNB is $\langle n\rangle=k(r-\Delta)$. The width squared
of the multiplicity distribution is $w\langle n\rangle\equiv
\langle n^2\rangle-\langle n\rangle^2=\langle n\rangle(1+r+\Delta)$.
With the MNB distributions for multiplicities of kaons and pions, the
distribution of event-by-event $K/\pi$ ratio is shown to be also
a Gaussian, similar to but broader than that given in Fig. 1.
Our calculations show that the mean value and the width of $K/\pi$
ratio distribution have weak dependence on the choice of $k$.
Thus, the parameter $k$ is chosen to be 200 in the following. In the
calculation of the dependences of the mean value and width of the $K/\pi$
ratio distribution on the pion
mean multiplicity, there are two free parameters $w_\pi$ and
$w_k=(\langle k^2\rangle-\langle k\rangle^2)/\langle k\rangle$
which represents the relative width of the distribution of kaon multiplicity.
Then, $R$ can be calculated from Eq. (\ref{ratio}) with $G_{\rm MNB}$
in place of $G_\Pi$. The width can also be calculated as
\begin{equation}
{\sigma\over R_e}=\sqrt{
{\left(1+{w_k\over R_g\langle\pi\rangle}\right)
\int_0^1dx\ {G_{\rm MNB}(x)-G_{\rm MNB}(0)\over x}\ln{1\over x}
\over \left[\int_0^1dx\ {G_{\rm MNB}(x)-G_{\rm MNB}(0)
\over x}\right]^2 }-1}\quad,
\label{width-nb}
\end{equation}

\noindent
We do not write $w_k$ in Eq. (\ref{width}) as it is simply 1, because there
the kaon multiplicity distribution is assumed to be a Poissonian.
Data from NA49 experiments show that $(\langle N^2\rangle-\langle
N\rangle^2)/\langle N\rangle$ is about 2.0 for all charged particles.
Considering the fact that most (about 90\%) charged particles
are pions, one can fix approximately $w_\pi=(\langle\pi^2\rangle-
\langle\pi\rangle^2)/\langle\pi\rangle=2.0$ for comparison with
the experiment. Numerical calculations show that $\sigma/R_e$
has a quite weak dependence on $w_\pi$. From the MNB distribution one can
get the two parameters in Eq. (\ref{mnbd}) for pion multiplicity distribution
as $r=(w_\pi-1+\langle\pi\rangle/k)/2$, $\Delta=(w_\pi-1-\langle\pi
\rangle/k)/2$ with $k$ preset to be 200. Dependence of $\sigma/R_e$ on the
global ratio $R_g$ can be expected from Eq. (\ref{width-nb}) in the same way
as from Eq. (\ref{width}), i.e., the larger $R_g$, the smaller $\sigma/R_e$
for fixed $\langle\pi \rangle$ and other parameters. It is shown that
$\sigma/R_e$ increases considerably with $w_k$. As argued in Ref. \cite{jeon},
the main contribution to the fluctuations of the $K/\pi$ ratio comes from the
fluctuation of multiplicity of kaons which are the particle species with fewer
mean multiplicity. Normally, the relative multiplicity fluctuation for
such particle species is larger. So in following
calculations, three values of $w_K=2.0, 2.5, 3.0$ are chosen to present
the dependence of $\sigma/R_e$ on the width of kaon multiplicity distribution.
Because the mean ratio is independent of $w_K$, the calculated $R$ are
the same and close to 1.0 within 2.5\% with the three choices of $w_K$,
and are given also in Fig. 2. However, $\sigma/R_e$ increases
considerably with $w_K$, as shown in Fig. 3. In those analyses in
\cite{gr97,fs99,croland} $\sigma/R_e$ from the mixed events is about
23\% with $\langle \pi\rangle$=270.13. Then from Fig. 3 one can see
that the calculation from the MNB distribution can give value of $\sigma/R$
consistent with the experimental result from the mixed events when
 $w_k$ is suitably chosen (less than 2.5).

However, the value of $w_K$ in real experiment is unknown yet and
may be different from 2.5 significantly. Thus, the calculated
result from pure statistical consideration cannot at present be compared
directly with that from the mixed events out of experimental data. The possible
difference between them may have deep physical implications. This may show
that there exist some other physical effects in the event-by-event $K/\pi$
distribution, which might be the correlation between the productions of kaons
and pions. Such effects should be studied theoretically and experimentally.

As a summary, the statistical influence on the $K/\pi$ ratio is studied
in an event-by-event way. It is shown that the distribution of the ratio
in the pure statistical case is Gaussian with mean and width depending
on the mean multiplicities of kaons and pions. It is interesting to further
study the effect on the ratio from correlations between the productions
of kaons and pions and the influence of other dynamical contributions.

This work was supported in part by NNSF in China.
C.B.Y would like to thank the Alexander von Humboldt Foundation
for the Research Fellowship granted to him.

\vskip 2cm

\begin{figure}[hb]
\centering
\includegraphics[width=0.5\textwidth]{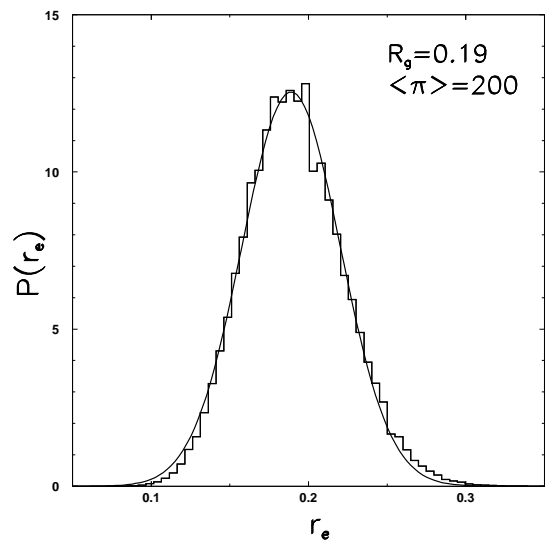}
\caption{\ The distribution of $K/\pi$ ratio from Monte Carlo
simulation for the independent production of kaons and pions with
the global $K/\pi$ ratio $R_g=0.19$ and mean pion multiplicity
$\langle\pi\rangle=200$. Both the distributions for kaons and pions
are assumed to be Poissonian. The solid curve is a Gaussian fit to the
distribution.}
\end{figure}

\begin{figure}
\centering
\includegraphics[width=0.5\textwidth]{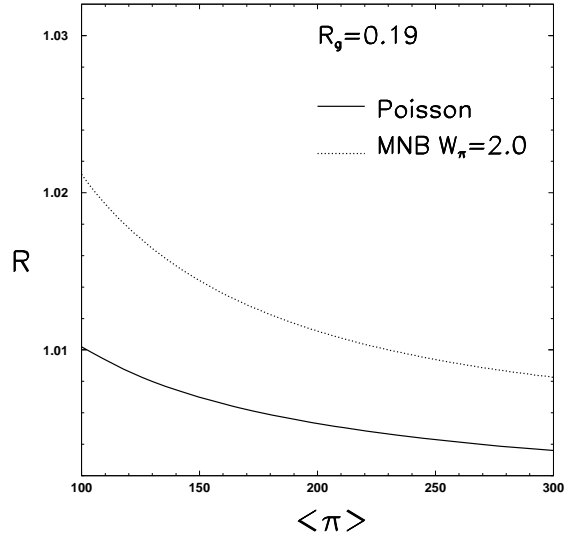}
\caption{\ The mean $K/\pi$ ratio from an event-by-event analysis
relative to $R_g$ as a function of $\langle\pi\rangle$ for the cases
with the Poisson and the MNB distributions for pion multiplicity.}
\end{figure}

\begin{figure}
\centering
\includegraphics[width=0.5\textwidth]{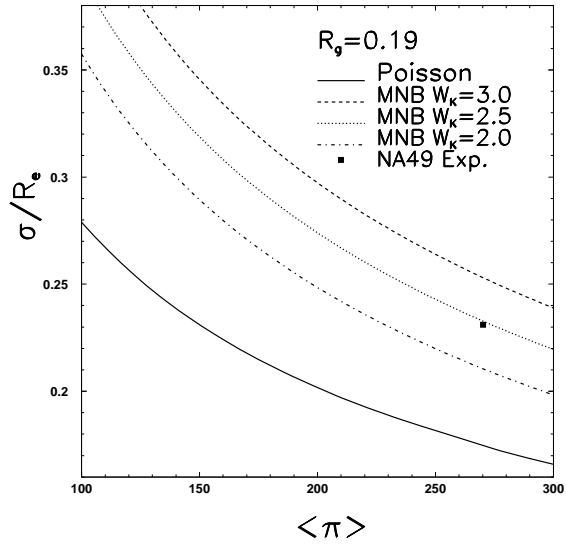}
\caption{\ The relative variance of the distribution of $K/\pi$ ratio
as a function of $\langle\pi\rangle$ for given $R_g=0.19$ for the cases
with the Poisson and the MNB distributions of the multiplicities of kaons and
pions. The experimental point from \cite{croland} is given by a solid square.}
\end{figure}
\end{document}